\documentclass[12pt]{emulateapj}
\usepackage{rotate}

\usepackage{natbib}
\bibpunct{(}{)}{;}{a}{}{,}

\def\XMM{{\em XMM--Newton}}
\def\pn{{\em pn}}
\def\BD{BD\,+37$^\circ$\,442}   

\def\HD{HD\,49798}
\def\approxgt{\mathrel{\hbox{\rlap{\lower.55ex \hbox {$\sim$}}
        \kern-.3em \raise.4ex \hbox{$>$}}}}
\def\approxlt{\mathrel{\hbox{\rlap{\lower.55ex \hbox {$\sim$}}
        \kern-.3em \raise.4ex \hbox{$<$}}}}

\def\ltsima{$\; \buildrel < \over \sim \;$}
\def\lsim{\lower.5ex\hbox{\ltsima}}
\def\gtsima{$\; \buildrel > \over \sim \;$}
\def\gsim{\lower.5ex\hbox{\gtsima}}

\shorttitle{Discovery of a compact companion to \BD}
\shortauthors{La Palombara et al.}

\begin{document}

\title{Discovery of a compact companion to the hot subdwarf star \BD}

\author{Nicola La Palombara, Sandro Mereghetti}
\affil{INAF - IASF Milano, via Bassini 15, I--20133 Milano, Italy}
\email{nicola@iasf-milano.inaf.it, sandro@iasf-milano.inaf.it}
\author{Andrea Tiengo}
\affil{IUSS-Istituto Universitario di Studi Superiori, viale Lungo Ticino Sforza
56, I--27100 Pavia, Italy}
\email{andrea.tiengo@iusspavia.it}
\author{Paolo Esposito}
\affil{INAF, Osservatorio Astronomico di Cagliari, localit\`a Poggio dei Pini,
strada 54, I--09012, Capoterra, Italy}
\email{paoloesp@oa-cagliari.inaf.it}

\begin{abstract}
We report the results of the first X--ray observation of the luminous and helium--rich O--type subdwarf \BD\ carried out with the \XMM\ satellite in August 2011.
X--ray emission is detected with a flux of about 3$\times10^{-14}$ erg cm$^{-2}$ s$^{-1}$ (0.2--1 keV) and a very soft spectrum, well fit by the sum of a blackbody with temperature $kT_{\rm BB} = 45^{+11}_{-9}$ eV, and a power law with a poorly constrained photon index. 
Significant pulsations with a period of 19.2 s are detected, indicating that the X--ray emission
originates in a white dwarf or neutron star companion, most likely powered by accretion from the wind of \BD. 
\end{abstract}


\section{Introduction}\label{introduction}

Hot subdwarfs are evolved, low--mass stars that have lost most of their
hydrogen envelope and are now in the stage of helium--core burning
\citep[see][ for a review]{heb09}. From the spectroscopic point of view we
distinguish the cooler B--type subdwarf (sdB) stars, with effective temperature 
$T_{\rm eff} <$ 40,000 K, and the hotter O--type (sdO) stars, with $T_{\rm eff} >$ 40,000 K \citep{hir08}. Most sdB stars are helium poor and display no or
only weak helium lines, while most sdO stars are helium rich and show He\,{\sc i} and
He\,{\sc ii} lines. While the sdBs form a homogenous group, the sdOs show a wide spread
in temperature ($T_{\rm eff}$ = 40,000--100,000 K), gravity (log($g$) = 4--6.5)
and helium abundance \citep{heb92,heb06}. Historically, sdO stars were
divided into `luminous' and `compact' sdOs, depending on their  
value of log($g$).

One possible mechanism responsible for the loss of the massive
hydrogen envelopes of hot subdwarfs is mass transfer in a binary. Indeed, there
is substantial evidence that many hot subdwarfs are in close binary systems
\citep{max01,nap04,cop11}. In particular, a large fraction of binaries is found among
sdB stars. Models of binary evolution \citep{han02} predict that many of the subdwarf 
companions should be white
dwarfs (WD). Indeed a few sdB+WD are known \citep{koe98,max00,ede05}, and recent optical
studies of single--lined spectroscopic binaries yielded several new candidates \citep{gei10}. 
The presence of a compact companion (a WD, neutron star (NS), or black hole) could be revealed
by the detection of X--ray emission powered by accretion, if the subdwarf
mass--donor can provide a sufficiently high accretion rate. 
Short Swift/XRT observations of candidate sdB+WD/NS binaries, selected from the \citet{gei10} 
sample, gave X--ray luminosity upper limits of $\sim10^{30}-10^{31}$ erg s$^{-1}$ \citep{mer11a}. 
These limits confirm that sdB stars have rather weak stellar winds,  
unable to provide enough accretion rate. 
 
Although the fraction of binaries among sdO stars is
not as high as for the sdBs, the prospects to find X--ray emitting
companions are more promising for these stars. In fact, at least a few
luminous sdO stars show evidence for stellar winds with mass loss rate $\dot M \sim
10^{-7}-10^{-10} M_{\odot}$ yr$^{-1}$ \citep{jef10}. Among these, the
bright star \HD\ is known as a soft X--ray source, with a strong periodic
modulation at 13.2 s, since more than 15 years \citep{isr97}. More recent
X--ray observations with \XMM\ showed that the compact companion of \HD\ is a
massive (1.28 $\pm$ 0.05 $M_{\odot}$) white dwarf \citep{mer09}. To our
knowledge, no other X--ray detections of sdO stars have been reported up to now.

Prompted by our findings on \HD, we carried out an X--ray observation of
another bright sdO star. Our target is \BD\ \citep{reb66}, a luminous
sdO which is often referred to as an `extreme helium
star' because of its complete lack of hydrogen (while \HD\ has about 20 \% (by mass) of hydrogen): it shows
evidence for significant mass loss rate
\citep{jef10} and stellar parameters similar to those of \HD\
(see Tab.~\ref{parameters}), but with the notable difference that,
up to now, no evidence of binary nature was reported.
Here we present the results of the first X--ray observations of \BD, which indicate
the presence of a compact companion, either a WD or a NS.

\begin{table*}[htbp]
\caption{Main parameters of the sdO stars \BD\ and \HD.}\label{parameters}
\begin{center}
\begin{tabular}{lccccc} \hline \hline
Parameter				& Symbol	& \multicolumn{2}{c}{\BD}			& \multicolumn{2}{c}{\HD}			\\
					&		& Value				& Reference	& Value				& Reference	\\ \hline
Mass ($M_{\odot}$)			& $M$		& 0.9				& 1		& 1.50 $\pm$ 0.05		& 5		\\
Radius ($R_{\odot}$)			& $R$		& 1.6				& 1		& 1.45 $\pm$ 0.25		& 6		\\
Surface gravity				& log $g$	& 4.0 $\pm$ 0.25		& 2		& 4.35				& 7		\\
Luminosity ($L_{\odot}$)		& $L$		& 25,000			& 3		& 14,000			& 7		\\
Effective Temperature (K)		& $T_{\rm eff}$	& 48,000 K			& 3		& 46,500			& 7		\\
Magnitudes				& $U,B,V$	& 8.57, 9.73, 10.01		& 4		& 6.76, 8.02, 8.29		& 8		\\
Distance (kpc)				& $d$		& 2$^{+0.9}_{-0.6}$		& 2		& 0.65 $\pm$ 0.1		& 6		\\
Terminal wind velocity (km s$^{-1}$)	& $v_{\infty}$	& 2,000				& 3		& 1,350				& 9		\\
Mass--loss rate ($M_{\odot}$ yr$^{-1}$)	& $\dot M$	& 10$^{-8.5}$			& 3		& 10$^{-8.5}$			& 7		\\ \hline
\end{tabular}
\end{center}
\begin{small}
References: 1 - \citealt{hus87}; 2 - \citealt{Bauer&Husfeld95}; 3 -
\citealt{jef10}; 4 - \citealt{Landolt73}; 5 - \citealt{mer09}; 6 -
\citealt{kud78}; 7 - \citealt{ham10}; 8 - \citealt{lan07}; 9
- \citealt{ham81}
\end{small}
\end{table*}

\section{Observations and data analysis}\label{data}

\BD~was observed with \XMM\ on 2011 August 21, starting at 23:35:19 UT (MJD =
55794.983). The three EPIC cameras, i.e.~one \pn\ \citep{str01} and two MOS
\citep{tur01}, were operated in \textit{full frame} mode, with time
resolution of 73 ms for the \pn\ and of 2.6 s for the two MOS cameras;
the effective source exposure time was, respectively, of $\sim$ 33 ks and $\sim$ 28 ks. 
For all cameras the medium thickness filter was used.

We used version 11.0 of the \XMM~{\em Science Analysis System} (\textit{SAS})
to process the event files. 
The observation was affected by high instrumental background only for a short 
time interval of $\sim$ 1.2 ks, which we removed for the spectral analysis.
We selected only the events 
with pattern in the range 0--4 (i.e.~mono-- and
bi--pixel events) for the \pn\ camera and 0--12 (i.e.~from 1 to 4 pixel events)
for the two MOS.

The images obtained with the three cameras clearly show that \BD\ is emitting in the
X--ray range: 
in fact a source is significantly detected at the coordinates R.A. = 01$^h$
58$^m$
33.4$^s$, Dec. = +38$^\circ$ 34$'$ 22.0$''$, which differ by only 1.8$''$ from
the 
position of \BD. This difference is consistent with the $\sim$ 2$''$ r.m.s. astrometric
accuracy of 
\XMM\  \footnote{http://xmm2.esac.esa.int/docs/documents/CAL-TN-0018.ps.gz}. 
The source net count rate in the 0.15--2 keV range is (1.62 $\pm$ 0.08)$\times$10$^{-2}$ cts s$^{-1}$ and (1.9 $\pm$ 0.2)$\times$10$^{-3}$ cts
s$^{-1}$ for the \pn\ and each of the two MOS, respectively.
\BD\ is not detected above $\simeq$ 2 keV, suggesting a soft spectrum, as confirmed by the spectral analysis presented below.

\begin{figure}[h]
\centering
\resizebox{\hsize}{!}{\includegraphics[angle=+90,clip=true]{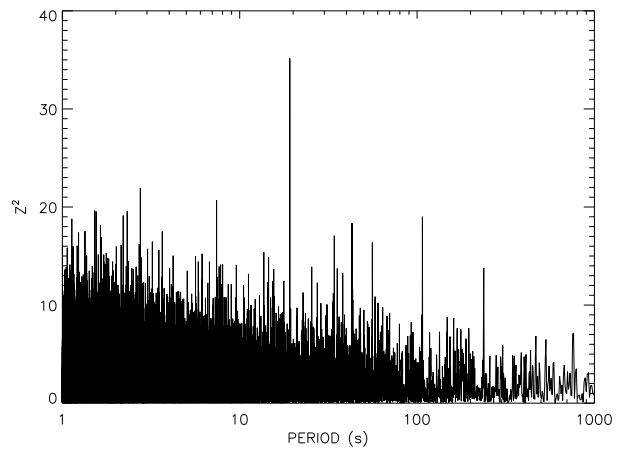}}
\caption{Distribution of the $Z^{2}$ statistics versus trial period for the \BD\ events in the 0.15--2 keV energy range.}
\label{period}
\end{figure}
 
\begin{figure}[h]
\centering
\resizebox{\hsize}{!}{\includegraphics[angle=-90,clip=true]{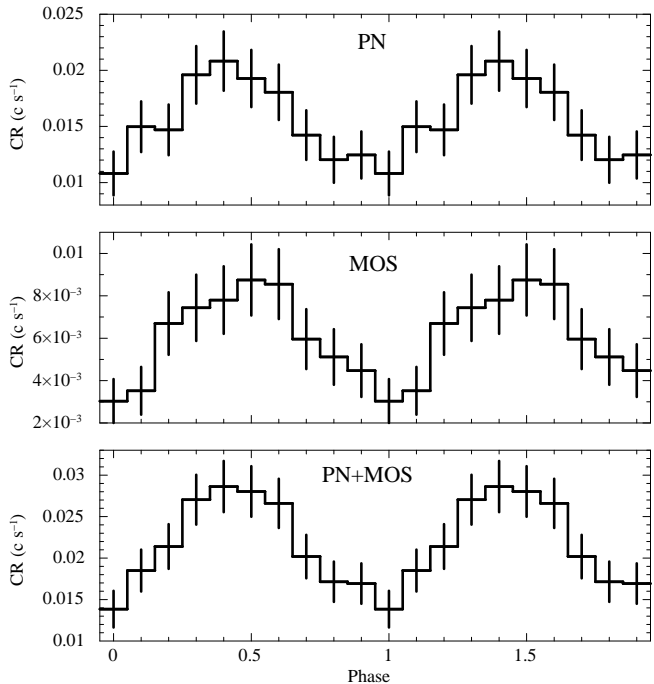}}
\caption{Background--subtracted light curves of \BD\ in the energy range 
0.15--2 keV, folded at the best--fit period \textit{P} =
19.156 s. From top to bottom: pn data, sum of MOS1 and MOS2 data, total data (pn+MOS1+MOS2).}
\label{flc}
\end{figure}

For the timing analysis, we used the data of the whole observation and 
the three EPIC cameras, extracting the events from a circular region with
radius 15$''$ centered at the source position. 
We converted the arrival times to the solar system
barycenter, then we
combined the three datasets in a single event list.
The background--subtracted light curve of \BD\ 
does not show variability on time scales from hundreds of seconds to the observation length.
In order to look for periodicities, we examined 65,232 statistically
independent periods in the range 1--10,000 s applying the phase--distribution test
\citep{buc83} to the counts in the 0.15--2 keV energy range (685 counts).
The resulting distribution of $Z^{2}$ statistics versus trial period is shown in
Fig~\ref{period}. A peak value of $Z^{2}$ = 35.15 is found at the period $P$ = 19.16 s. The corresponding probability of chance occurrence, after taking into account the
number of trials, is 1.5 $\times$ 10$^{-3}$. 
As shown by the folded and background--subtracted light curves of Fig.~\ref{flc},
the periodicity is visible also in the
separate data sets, i.e. \pn\ and MOS1+MOS2,
which give consistent phase distributions and pulsed fractions.
The refined period value, obtained through pulse
phase fitting, is $P$ = 19.156 $\pm$ 0.001 s. 
The pulse profile shows a single broad peak with a 0.15--2 keV pulsed
fraction of 31 $\pm$ 4 \%  (amplitude of a sinusoid fitted to the folded light curve
divided by the average flux).
No significant differences are seen in the light curve shape and pulsed fraction in 
different energy ranges. We also performed a phase--resolved spectral analysis, as described below, for pairs of spectra corresponding
to the pulse maximum/minimum or rise/decrease phase intervals, without
finding significant differences in the spectral parameters.

Due to the limited signal--to--noise ratio of the MOS  data, for the
spectral analysis we considered only the \pn\ events. We generated the
applicable response matrices and ancillary files using the \textit{SAS} tasks
\textit{rmfgen} and \textit{arfgen}. To ensure the applicability of the
$\chi^{2}$ statistics, the spectrum was rebinned with a minimum of 30 counts per
bin and fitted using \textit{XSPEC} (V 12.7.0). We only used the energy range 0.2--1 keV since above 1 keV the
background dominates and the source flux is negligible. In the following, all the spectral uncertainties and upper
limits are given at the 90 \% confidence level for one interesting parameter,
and we assume a source distance of 2 kpc \citep{hus87}.

\begin{figure}[h]
\centering
\resizebox{\hsize}{!}{\includegraphics[angle=-90,clip=true]{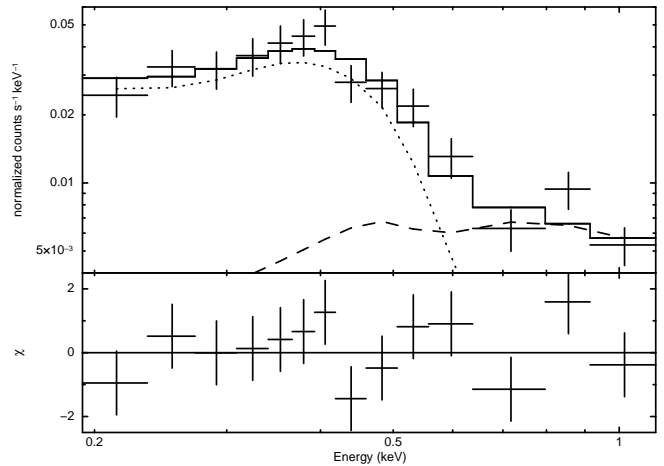}}
\caption{\textit{Top panel}: \pn\ spectrum of \BD\ with the best--fit power--law
(\textit{dashed line}) plus blackbody (\textit{dotted line}) model.
\textit{Bottom panel}: residuals (in units of $\sigma$) between data and
model.}
\label{spectrum}
\end{figure}

The source spectrum is very soft: a fit with an absorbed power law gives an
unphysically large photon index ($\Gamma \sim 5$), while a blackbody gives a temperature $kT_{\rm BB}$ = 93 eV. However, these single--component models give unacceptable fits
($\chi^{2}_{\nu} >$ 2). A model consisting of an
absorbed power law plus blackbody provides a significant
improvement ($\chi^{2}_{\nu}$ = 1.24 for 9 degrees of freedom, Fig.\ref{spectrum}), 
although the power--law photon index $\Gamma$ is poorly
constrained in the range between -1 and 4. Keeping $\Gamma$ fixed at its best
fit value (2.25), the values of the other spectral parameters are: column density
$N_{\rm H} = (1.4^{+0.7}_{-0.6} )\times 10^{21}$ cm$^{-2}$, blackbody temperature $kT_{\rm BB} =
45^{+11}_{-9}$ eV,
and emitting radius $R_{\rm BB} = 39^{+162}_{-28}$ km. The absorbed flux in
the energy range 0.2--1 keV is $f_{\rm abs,X} = (2.6 \pm 0.3) \times 10^{-14}$
erg cm$^{-2}$ s$^{-1}$, of which 28 \% can be ascribed to the power--law component.
The bolometric flux of the blackbody component, corrected for the absorption, is $1.7\times 10^{-12}$
erg cm$^{-2}$ s$^{-1}$, corresponding to a luminosity of 7.8$\times 10^{32}$ erg s$^{-1}$ (for a source distance $d$ = 2 kpc).
This value is subject to a large uncertainty because the 
value of $kT_{\rm BB}$ is only poorly constrained and depends on the slope of the power--law component. Therefore,
in order to assess in a conservative way the uncertainty on the source luminosity, we repeated the
spectral fits fixing the photon index at the values $\Gamma$ = 1 and $\Gamma$ = 3. 
In this way, we found acceptable values of $kT_{\rm BB}$ in the range 30--68 eV (90 \% c.l.), corresponding to bolometric blackbody luminosities between  
6$\times 10^{31}$ and 1.5$\times 10^{35}$ erg s$^{-1}$ (where the highest luminosity is obtained for the lowest temperature). The 0.2--10 keV luminosity of the power--law component is in the range (1.8--3.2)$\times 10^{31}$ erg s$^{-1}$.


\section{Discussion}

Our \XMM\ observation of \BD\ has provided the first detection of X--ray emission
from this O--type subdwarf.
Shocks in the radiation--driven stellar winds  
of main sequence, giant, and supergiant early type stars give rise to 
soft X--rays, with a typical ratio of X--ray to optical flux of the order 
of log($f_{\rm 0.5-10~keV}^{\rm unabs}/f_{\rm bol}$) = -6.45 $\pm$ 0.51 \citep{naz09}.
It is not known if this relation holds also for low luminosity stars. 
The only other sdO star possibly detected in X--rays is \HD, which is in a binary
system with an X--ray emitting white dwarf. In this system, the  X--ray
flux of $f_{\rm X} = 4.3 \times10^{-14}$
erg cm$^{-2}$ s$^{-1}$, visible when the accreting WD companion is
in eclipse, corresponds to log($f_{\rm 0.5-10~keV}^{\rm unabs}/f_{\rm bol}$) = -7.4
and could be due to emission from the sdO \citep{mer11b}.  
For \BD\ we measure log($f_{\rm 0.5-10~keV}^{\rm unabs}/f_{\rm bol}$) = -6.63, 
consistent with the above values, but the significant periodic modulation at 19.156 s clearly indicates
that most of, or possibly all, the detected flux originates in a compact companion star,
rather than in \BD\ itself. The periodic modulation can be explained equally well as the spin period of a NS or of a WD. 

\begin{figure}[h]
\centering
\resizebox{\hsize}{!}{\includegraphics[angle=0,clip=true]{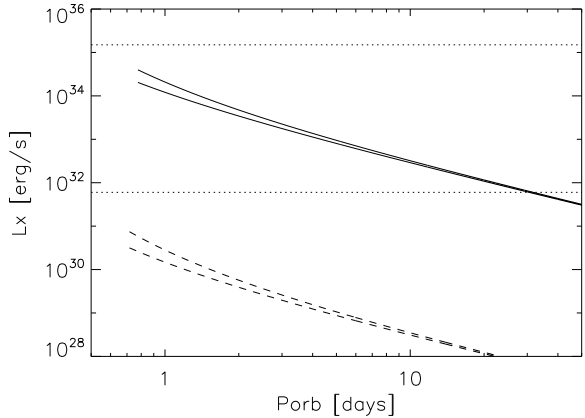}}
\caption{Estimated wind--accretion X--ray luminosity of \BD\ as a function of the binary period,
in the case of a WD (\textit{dashed lines}) and of a NS (\textit{solid lines})
The couples of lines refer to two values of the wind velocity index $\beta$ = 0.6 and $\beta$ = 1.
The horizontal dotted lines represent the minimum and maximum \textbf{of the estimated} luminosity of the blackbody component.}
\label{luminosita}
\end{figure}

Ultraviolet spectra of \BD, acquired with the IUE satellite, show N\,{\sc v} and C\,{\sc iv} resonance lines with P Cygni-like profiles, indicating the presence of a stellar wind \citep{ros84}. A more recent analysis of UV and optical spectra of \BD\ has been 
reported by \citet{jef10}, who derived a mass loss rate of 3$\times$10$^{-9}$ $M_{\odot}$  yr$^{-1}$ and
a wind terminal velocity of $v_{\infty}$ = 2,000 km s$^{-1}$. It is possible that part of the sdO stellar
wind is captured by its compact companion, giving rise to accretion powered X--ray emission.
The density and velocity of the wind material at the position of the compact object   
can be computed assuming a canonical wind
velocity law with radial dependence $v(R) = v_{\infty} (1-1/R)^{\beta}$, where $R$ is the radial distance in units of stellar radii and the
index $\beta$ is typically in the range 0.6--1. Assuming for simplicity Bondi--Hoyle accretion, and relating
the orbital separation  to the period of a circular orbit $P_{\rm orb}$ through Kepler's law, we can derive
the expected accretion luminosity as a function of $P_{\rm orb}$ shown in Fig.~\ref{luminosita}.
The curves refer to a NS of 1.4 $M_{\odot}$ and 10 km radius, and to a WD of 0.6 $M_{\odot}$  
and 10,000 km radius; in both cases we used for \BD\ a mass of 0.9 $M_{\odot}$ \citep{hus87}.
We have also indicated in the figure the allowed range of luminosity of the blackbody component
(for simplicity we do not consider the additional 30 \% contribution to the total luminosity coming from the power--law component).

Although the observed luminosity is only poorly constrained due to the large uncertainties in the spectral parameters, it is clearly consistent 
with that expected from a NS orbiting \BD\ with a period between about one and several days. 
The blackbody emitting radius derived from the best fit, 
$R_{\rm BB} = 39^{+162}_{-28}$ km (for $d$ = 2 kpc), is only marginally consistent with a neutron star. 
However, this parameter is strongly correlated with the poorly constrained slope
of the power law spectral component and an acceptable fit can be obtained, e.g., with
$\Gamma$ = 1.5, $kT_{\rm BB}\sim$ 58 eV, and $R_{\rm BB}\sim$ 10 km.
Alternatively, the 
accreting companion could be a WD, if the sdO extends to (or close to) the
Roche--lobe, thus yielding an accretion rate larger than what we computed assuming 
wind accretion (or in the unlikely case that the adopted distance
is largely overestimated).
The observed spectrum is much softer than the typical spectrum of neutron stars in classical X--ray binaries, 
which however have very different companion stars and accrete at higher rates. It should also
be considered that we can observe the low temperature emission from \BD\ (and \HD, see below) only because these sources are at
high Galactic latitude, where interstellar X-ray absorption is very small.

Up to now \BD\ was believed to be a single star: no evidence for a companion was seen in 
spectroscopic \citep{fay73,kau80,dwo82} or photometric data \citep{Landolt68,Landolt73}, and infrared observations did
not show any excess emission \citep{Thejll+95}. On the other hand, some of the very few published radial 
velocity measurements are inconsistent with a single value, hinting to a possible binary motion. A radial velocity $V_{\rm r}$ = --156.4 $\pm$ 1.1 km s$^{-1}$
is given in the original discovery paper \citep{reb66}, while \citet{dri87} found $V_{\rm r}$ = --94 $\pm$ 1 km s$^{-1}$.
New accurate radial measurements are needed to determine the parameters of this binary.

The X--ray properties of \BD\ are very similar to those of the only other X--ray source associated with an sdO star: \HD. In this binary a massive white dwarf 
accretes matter from the stellar wind of its subdwarf companion \citep{mer09}. 
Also in the  case of \HD\ the X--ray emission is pulsed, with a period (13.2 s)
similar to that we have discovered in \BD, and the soft spectrum
is well described by the sum of a blackbody with $kT_{\rm BB}$ = 39 eV  
and a power law with $\Gamma$ = 1.6 
\citep{mer11b}. \HD\ has an orbital period of 1.55 days, first discovered
through radial velocity measurements of the sdO optical emission \citep{tha70}, 
and an X--ray luminosity of 10$^{32}$ erg s$^{-1}$ (at $d$ = 650 pc).

The fact that the only two sdO stars observed with sensitive X--ray 
instruments turned out to be binaries with a compact companion is noteworthy.
Although it is premature to draw conclusions from such small numbers, 
this might suggest that the fraction of sdO binaries is larger than 
currently believed on the basis of optical observations. 
Note that, contrary to late type companions, which
can give a detectable contribution in the spectral and photometric
data, WDs and NSs are too faint and
completely outshined in the optical/UV by the sdO emission.
Both \BD\ and \HD\ belong to the subclass of luminous, 
He-rich sdOs, whose origin and evolutionary link with other classes of stars
is still unclear \citep{nap08,jus11}. In this respect, X--ray observations might provide 
important information to complement optical/UV data. Besides the potential of
discovering other sdO binaries through the detection of pulsations, the high sensitivity of current satellites like \XMM\ and Chandra can probe the X--ray emission
of single sdOs, and test whether their wind emission scales as in early type stars of higher luminosity.



\acknowledgements

We wish to thank U. Heber and S. Geier for useful discussions.
This work is based on observations obtained with \XMM, an ESA science mission
with instruments and contributions directly funded by ESA Member States and NASA.
We acknowledge financial contributions by the Italian Space Agency through ASI/INAF
agreements I/009/10/0 and I/032/10/0 for, respectively, the data analysis and the
\XMM\ operations. P.E. acknowledges financial support from the Autonomous Region
of Sardinia through a research grant under the program PO Sardegna FSE 2007--2013,
L.R. 7/2007 `Promoting scientific research and innovation technology in
Sardinia'.

.

\end{document}